# Inverse design for scalable photonic systems

*Louise Schul[1], Sydney Mason[1], Sungjun Eun[1], Geun Ho Ahn[1] and Jelena Vučković[1]**

[1] Department of Electrical Engineering and Ginzton Laboratory, Stanford University, Stanford, CA 94305-4088, USA.

*e-mail: jela@stanford.edu

**Abstract |** Over the past two decades, photonic inverse design has emerged as a powerful approach to implement photonic devices with improved performance, or realize new functionalities. While the efforts over the first decade focused on proof of concept devices designed and fabricated in university labs, the focus over the past 5-10 years has shifted towards implementation of scalable photonic systems. This article reviews this recent progress, challenges and new directions in photonics inverse design, thus providing a complementary and updated review of the field. We focus on large scale three dimensional photonic inverse design, including metasurfaces, translation of inverse design to commercial foundries and practical silicon photonics, application of photonic inverse design to different materials systems, wavelengths, and optical effects, and finally new directions such as inverse design of quantum systems.

## Introduction

Photonic inverse design is an approach in which photonics is designed by defining desired input to output optical field mapping and then by efficiently searching through the full parameter space in order to find the geometry that satisfies the desired functionality with any additional imposed constraints (fabrication, environment, bandwidth, efficiency, etc.)[1,2]. For nearly all photonics problems of interest, this search has to be performed in a three-dimensional (3D) parameter space in order to accurately capture and describe the device performance. Since the parameter space is enormous even for modest footprint devices, blind search is impossible, and physics guided search is crucial. For example, the use of 20 nm binary pixels for a 3D device with a linear dimension of 2 μm leads to $2^{1,000,000}$ possibilities - more than the number of atoms in the observable universe. Physics guided search means gradient descent towards a local optimum through this complex parameter space, by calculating spatial gradients to the figure of merit that is being optimized. Since gradient calculation is computationally very expensive, this process is done by adjoint optimization - mathematically very similar to back-propagation. Adjoint optimization computes full spatial gradients in devices with only two full-wave simulations per iterations: a forward simulation with the expected input excitation, and an adjoint (backward) simulation with a source constructed from the desired output, exciting the structure backwards.

Using a simple mathematical trick, structure gradients can be calculated from these two simulations, as shown in Figure 1, allowing us to update the structure in the way that would lead to the local optimum[1,2]. This process is typically repeated several hundreds of times, until a local optimum that satisfies requirements and constraints is found, requiring several hundreds of electromagnetic simulations to be done in series (two per step). Since electromagnetic simulations are generally slow and computationally expensive, especially for 3D devices, one of the main challenges and goals in photonic inverse design is the development of fast electromagnetic simulators, which are scalable to large 3D footprints[3]. Further complicating this process is the fact that for practical devices, optimization is constrained (binary structure, minimum feature sizes, large bandwidth etc.).

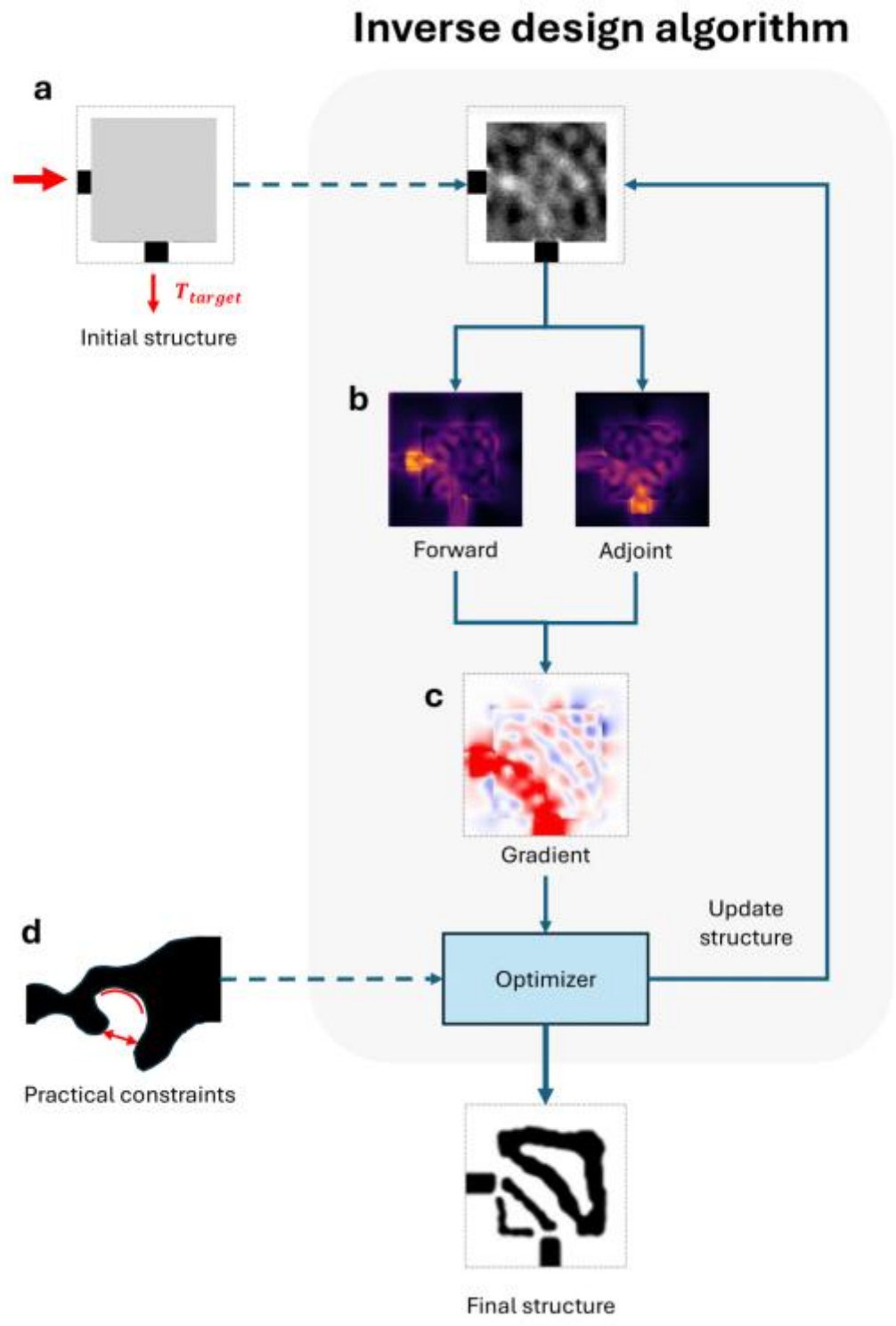

**Figure 1. Photonics inverse design of a waveguide bend via adjoint optimization**[2]
(**a**) The structure is initialized as a black box with fixed footprint and desired transmission efficiency. (**b**) The structure gradient is calculated from two electromagnetic simulations: forward and backward (adjoint). This process is similar to backpropagation in AI. (**c**) Gradient descent efficiently searches through the parameter space. (**d**) Practical devices require constraints such as binary structure and minimum feature sizes.

This article reviews recent progress, challenges and new directions in photonics inverse design, focusing on the recent developments since the previous extensive review of the field in 2018[1]. In particular, we focus on what is we believe are the main developments in recent years: large scale three dimensional photonic inverse design, including metasurfaces (Section 2), translation of inverse design to commercial foundries and practical silicon photonics (Section 3), application of photonic inverse design to different materials systems, wavelengths, and optical effects (including nonlinear)(Section 4), and finally new directions in photonic inverse design focusing in particular on nonlinear optical and quantum systems (Section 5). Figure 2 depicts an overview of these developments.

## Large-scale photonics inverse design

As electromagnetic solvers have advanced[3–5], inverse design has become pervasive in fields that are historically constrained by simulation size, such as free-space optics. Metasurfaces are typically made up of a periodic array of subwavelength-scale structures and can impart significant manipulation over all the properties of light at the nanoscale, providing new functionalities for free-space beam control and a more efficient and miniaturized solution to bulk optic counterparts. Enabled by larger simulation regions and creative approximation techniques, there has been a recent explosion of inverse design in the metasurface field. A variety of methods have been explored in recent years to exploit inverse design concepts and depart from the traditional manual parameter tuning and iterative brute force search design method that has been used for decades. We note that there are other pre-existing comprehensive reviews of inverse design metasurfaces[6,7]. For the purpose of categorizing inverse design regimes by simulation size, we define large-scale inverse design to be on the order of > 20 μm x 20 μm (or ~ 50λ/n) and larger in linear dimension), medium-scale around 10 - 20 μm and small-scale as < 10 μm. Due to the dominance of metasurfaces in existing work on large-scale inverse design, we mainly focus the following section on metasurface design approaches.

For the inverse design of metasurfaces, three main approaches have emerged. Firstly, the simulation of a single metaatom required in building a library of shapes is not computationally expensive. Yet, even with this pre-existing library the design of large area metasurfaces has been limited by the computational power required to calculate the desired phase profile. The state-of-the-art approach for the inverse design of a metasurface phase profile uses a combination of a fast approximate solver for far-field calculations and the adjoint method to design and experimentally demonstrate metaoptics with size on the order of 20,000λ[8] (Fig. 2k). In this approach, the authors use a training set of metaatoms and their simulated field response to predict the local field of an arbitrary metaatom. Designing a metasurface with multiple outputs proves to be computationally infeasible with classical approaches, but this can be circumvented with inverse design. For example, it is possible to encode 9 holograms into 1 metasurface and create 3D holograms using phase profile optimization[9] as well as use the adjoint gradient method to design an achromatic metalens doublet with a large NA, a metasurface doublet with 5 different wavelength-dependent holograms, and a multiple metasurface optical neural network for classifying digits[10]. This can be extended to more dimensions, enabling multi-channel imaging (depth/spectral/polarization) with an inverse-designed metasurface multiplexer, spatially separating wavelengths as narrow as 20 nm apart[11] (Fig. 2e). Other approaches overcome limitations in full-wave simulation capacity by using neural networks[12,13], genetic algorithms[14], and coupled-mode theory[15,16]. These methods have been used to design large area high-efficiency metasurfaces in optical and THz frequencies. Notably, in many cases using the adjoint method can result in the optimization being stuck in some non-optimal local minimum of the loss function as a result of the initial condition. To resolve this issue an approach coined "GLOnet-based optimization" can be utilized, which enables a global search of the design space[17].

The next approach to the inverse design of metasurfaces is topology optimization at the unit cell level. This method allows for the design of metaatoms with unintuitive geometries that when placed in a periodic array, perform precisely defined functions like beam deflection and focusing at GHz frequencies[18], polarization filters, half- and quarter-wave plates[19,20], producing arbitrary spectral responses[21], and solving integral equations[22]. Most metasurface design approaches ignore inter-metaatom coupling of dissimilar metaatom shapes, which can limit performance. In Dainese et al.[23], they develop a shape optimization method using the Fourier decomposition of the surface gradient to optimize the metaatoms of an existing library, allowing only for fabrication-friendly smooth variations to the geometries.

Finally, metasurface inverse design extends beyond phase profile optimization and unit cell design and into free-form optimization of the entire surface topology. These approaches span from 1D grating optimizations for large-area metasurfaces ($10^4$-$10^5\lambda$)[24] to the design of 3D metaoptics using full-wave FDTD simulations and the adjoint method[23], and specialized solvers for full metasurfaces such as T-matrix method[25]. The power of extending inverse design into multi-layer 3D structures has been demonstrated with 6-layer metaoptics that sort angular momentum, polarization, and spectrally filter in the mid-infrared[26] (Fig. 2h). Free-form metasurface inverse design has enabled new functionalities such as complementary scanning lenses for THz imaging[27] and tailored spatial and spectral responses that recreate the color perception of the human eye[28]. These demonstrations would previously have been unfeasible with the classical unit cell approach. Even with current state-of-the-art electromagnetic solvers, free-form 2D optimization of large-area metasurfaces remains a challenging task. In certain design scenarios, such as a lens, there are inherent symmetries which allows for the clever avoidance of optimizing the entire metasurface. Instead, the metalens can be broken into radial zones, allowing for the full wave simulation in cylindrical coordinates while exploiting a GPU-accelerated FDTD solver[29]. This technique facilitates the realization of a wide-aperture freeform 3D metalens (Fig. 2p)[29].

Additional applications, such as fiber couplers, can benefit from imposing radial symmetry to reduce the design region size for 3D structures. In order to improve the coupling efficiency between single-mode fibers and photonic wire bonds, gradient-based inverse design has been used to design 3D nanoprinted couplers[30]. This approach is used for couplers on the length scale of 25 μm and greater, achieving coupling efficiencies between the single mode fiber and photonic wire bond of > 90%. Large area inverse design has also been leveraged for on-chip applications that require longer propagation lengths such as mode sorting on a surface plasmon polariton platform, where a design region of 52 x 38 μm was used[31].

Efforts towards scaling inverse design structures have permeated medium-scale devices like fiber-to-chip couplers. Grating couplers were an excellent initial demonstration for inverse design with inverse design 1D gratings exceeding the performance of traditional grating couplers with uniform or apodized periodicity[32]. Since these preliminary demonstrations, design regions have been expanded to a 2D area on the order of 10 x 10 μm to 14 x 14 μm to accommodate telecom fiber modes with spot sizes of ~10 μm[33–35]. Coupling efficiencies of less than -3.4 dB with >60 nm 3-dB bandwidth have been realized in the O- and C-band of telecommunications[33]. Inverse design has been shown to be a suitable solution to grating couplers for vertically incident light which provide increased design difficulty. Commercial foundry-compliant multi-layer inverse design grating couplers have been demonstrated to achieve -4.7 dB coupling efficiency for vertically incident light[34]. This efficiency can be increased even further when a reflector is added below the grating coupler, enabling sub-dB coupling efficiency for a 14 x 14 μm coupler[35].

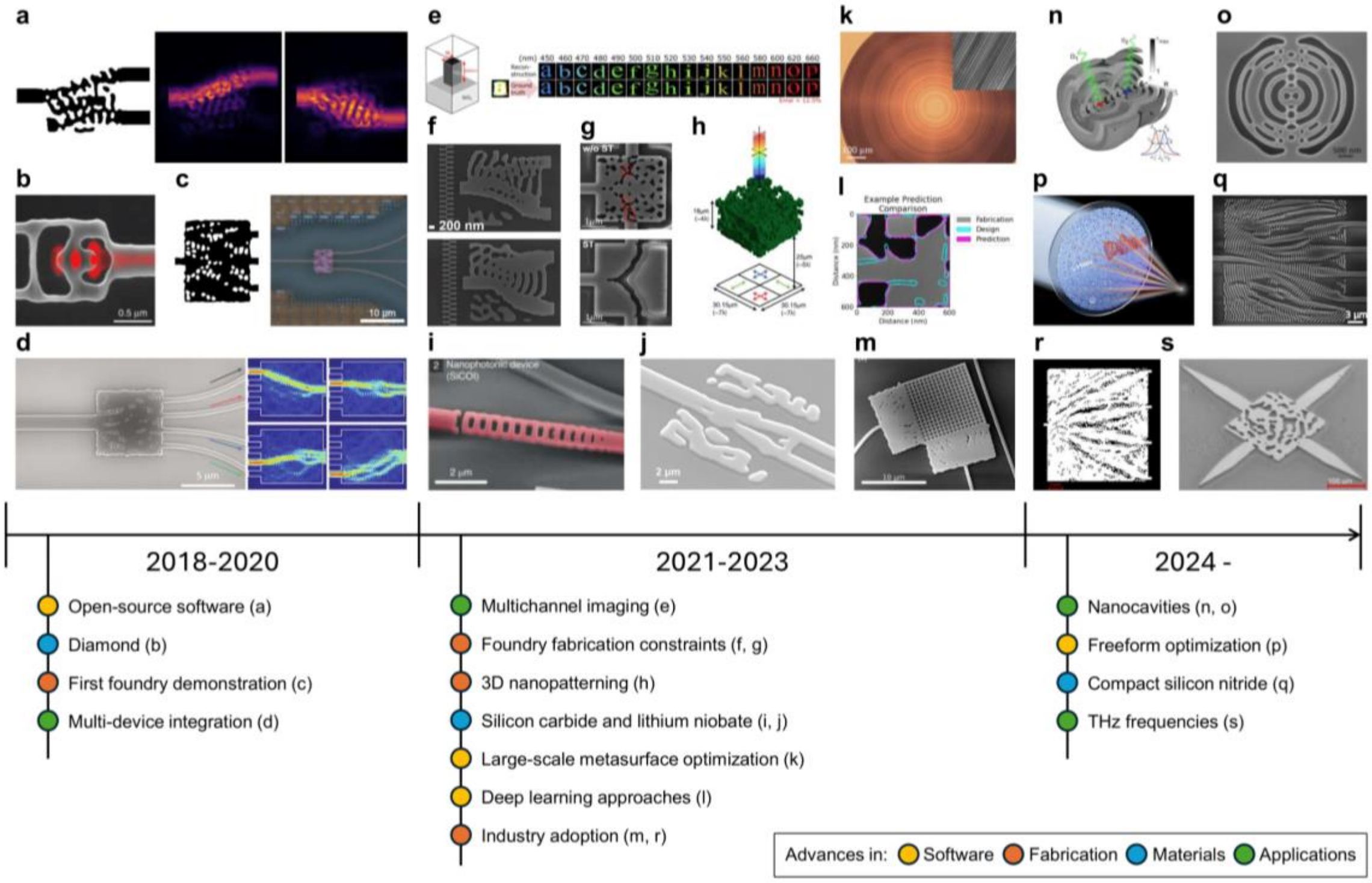

**Figure 2. Progression of inverse design 2018-2025**

(**a**) two-channel silicon-on-insulator WDM using SPINS from Su et al.[2](2019), (**b**) diamond vertical coupler from Dory et al.[77] (2019), (**c**) three-channel WDM fabricated in a commercial silicon foundry from Piggott et al.[36] (2020), (**d**) four-channel broadband low-crosstalk MDM from Yang et al.[52] (2022), (**e**) 16-channel metasurface spectral imager from Lin et al.[11] (2022), (**f**) 1310 nm/1550 nm wavelength duplexer optimized with energy constraint from Zhang et al.[39] (2022), (**g**) T-junction optimized with structure transformation from Chen et al.[45] (2023), (**h**) 3D-patterned multispectral and polarization sorting device from Roberts et al.[26] (2023), (**i**) silicon carbide reflector in Fabry-Perot cavity for nonlinear light generation from Yang et al.[83] (2023), (**j**) two-channel lithium niobate MDM from Shang et al.[73] (2023), (**k**) RGB-achromatic polarization-insensitive meta-lens from Li et al.[8] (2023), (**l**) deep-learning model to predict fabrication outcomes from Grinberg et al.[40] (2023), (**m**) compact dual-polarization silicon integrated couplers for multicore fibers from Ruiz et al.[50] (2021), (**n**) topology-optimized cavity hosting quantum emitter pair from Miguel-Torcal et al.[117] (2024), (**o**) InP nanocavity with extreme dielectric confinement from Xiong et al.[95] (2024), (**p**) wide-aperture freeform 3D metalens from Sun et al.[29] (2025), (**q**) four-channel compact WDM in silicon nitride from Ruiz et al.[63] (2025), (**r**) four-channel compact WDM designed at Google from Cheung et al.[49] (2024), (**s**) three-channel active WDM at THz frequencies from Digiorgio et al.[86] (2025)

Insets reproduced with permission from references [2, 77, 36, 52, 11, 39, 45, 26, 83, 73, 8, 40, 50, 117, 95, 29, 63, 49, 86]

## Inverse design translation to commercial foundries

Building on over a decade of proof of concept photonic inverse design demonstrations[1], its natural progression to high throughput fabrication and commercial semiconductor foundries has happened over the past 5-10 years. Commercial semiconductor manufacturing relies on photolithography processes with considerably lower resolutions than electron-beam lithography and ion beam machining[36]. To ensure manufacturability, foundries impose geometrical constraints including minimum linewidth, linespacing, curvature, area, and enclosed area. Designs must pass thousands of design rule checks (DRC) to be fabricated[37,38]. This process is illustrated in Figure 3.

Inverse-designed devices frequently include small features that violate DRC and are therefore difficult or impossible to fabricate[36,37,39–44]. Moreover, inverse-designed devices are highly sensitive to minor fabrication variations leading to performance degradation[38,39,43]. These challenges motivate inverse design frameworks that integrate manufacturability and robustness directly into the optimization process.

Strategies for robust DRC-compliant designs include 1) imposing analytic constraints during the optimization, 2) selecting a design parameterization that guarantees geometric feasibility, 3) integrating lithography models into the optimization process, and 4) employing machine learning techniques.

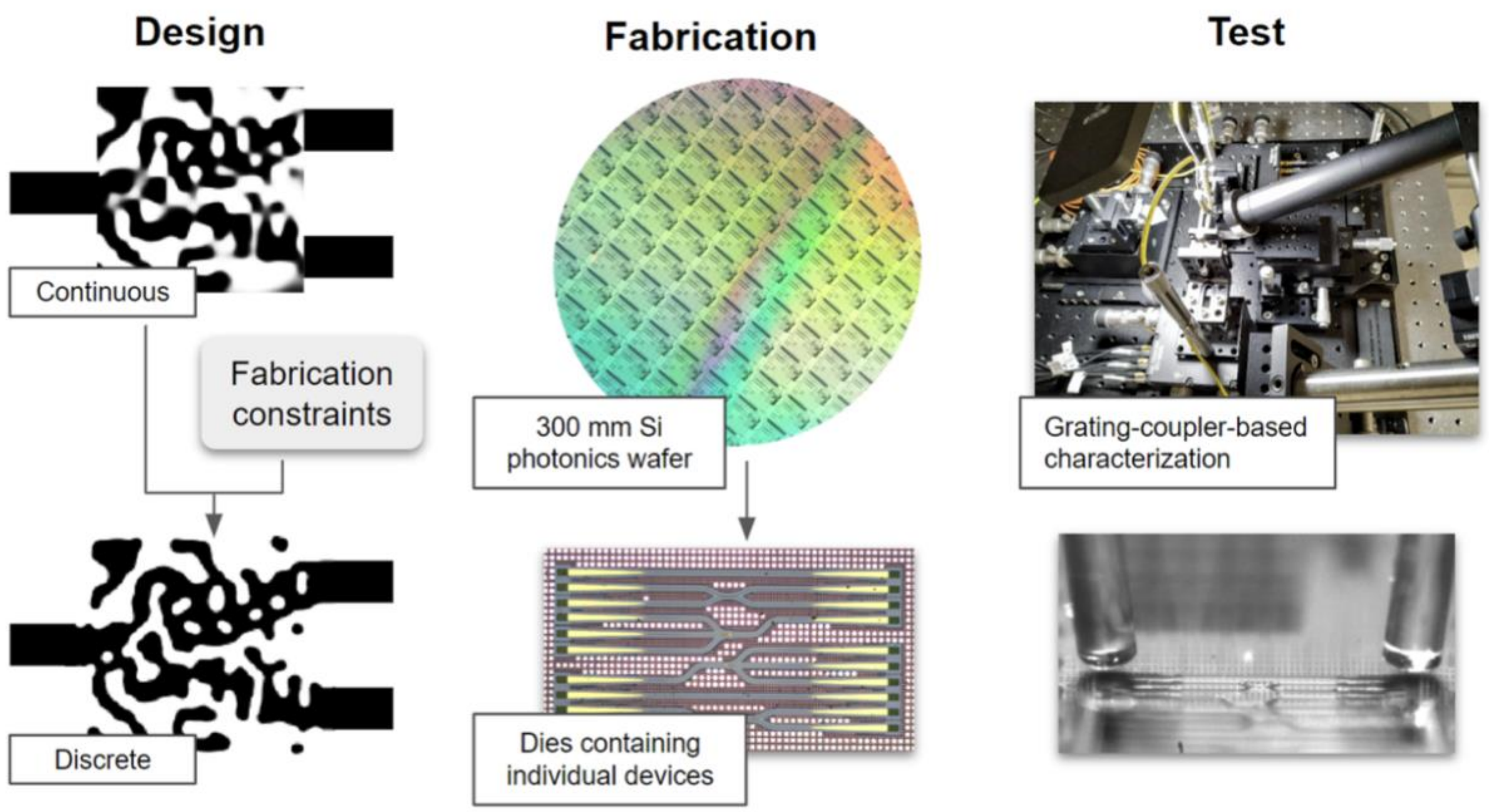

**Figure 3. Inverse design workflow for commercial silicon photonics foundries**
**Design:** fabrication constraints are applied in the discrete optimization stage. The final device design is then combined with components such as waveguides and grating couplers from the foundry's Process Design Kit (PDK) to complete the final mask pattern. **Fabrication:** AIM Photonics 3000 nm multiproject Si wafers are fabricated via water-immersion deep UV photolithography at the Albany NanoTech fabrication facility. **Test:** the wafer is diced and the devices are tested in a vertical transmission measurement setup. Reproduced with permission from Piggott et al.[36]

Imposing analytic constraints during optimization is a common strategy to guide the process towards manufacturable structures. The first successful demonstration of inverse design in a commercial silicon process incorporated heuristics for minimum feature gap and curvature into gradient-based optimization[36]. The authors fabricated a spatial mode multiplexer, a wavelength demultiplexer (Fig. 2c), a 50-50 directional coupler, and a 3-way power splitter and verified that all four designs are robust to fabrication errors, with ±0.6 dB device-to-device variability across three dies.

The introduction of additional 2D constraints on minimum area and enclosed area produced broadband waveguide bends, T-splitters, and reflectors complying with nine different foundry specifications[38]. As an alternative to geometric constraints, imposing an energy constraint can direct the optimization process to solutions that best contain the field inside silicon[39,40,44]. This method offers three advantages. 1) Naturally generates devices that are closer to binarized, 2) less energy interacts with boundaries and is therefore less affected with dimensional variations, and 3) naturally limits the generation of small features. When applied to 3D optimization for three types of devices (mode converter, O-band/C-band duplexer (Fig. 2f), and C-band three-channel WDM), the energy constraint improved binarization and reduced the presence of small features[39].

Alternatively, 'always-feasible' design methodologies select design parameterizations that guarantee manufacturability without the need for analytic constraints. Notably, topology optimization generates intermediate permittivity values that cannot be fabricated and thus requires a multi-stage optimization process. A typical workflow begins with initial optimization of a continuous permittivity distribution, followed by discretization and further optimization under fabrication constraints. When applied to a silicon-on-insulator two-channel WDM, the discrete optimization stage recovered the performance achieved in the continuous stage[2] (Fig. 2a). Several approaches have been proposed to produce manufacturable designs without postprocessing. Examples include an always-feasible framework that combines a conditional generator and straight-through gradient estimator to produce designs that are guaranteed to satisfy length-scale constraints,[37] a structure transformation technique that alternates between optimization steps and curvature-constrained structure adjustments to facilitate simpler topologies[45] (Fig. 2g), and incorporating a two-level hyperbolic tangent projection function into the optimization process[46]. Another approach seeds density-based topology optimization with a known functional geometry to guide towards ultra-compact, high-performance devices. When applied to a 2D modal multiplexer, 87% of devices conformed to foundry constraints, compared to only 13% of devices designed by conventional topology optimization[47]. A recent open-source software package consolidates previously-explored topology optimization techniques with a hybrid time/frequency-domain adjoint-variable formulation, thereby combining the broadband optimization of time-domain solvers with the flexibility and generalizability of frequency-domain solvers. This method achieved broadband performance from large-scale devices including a 3D polarization splitter and high-NA cylindrical metalens[48]. Departing from topology optimization, a tile-based framework combines direct binary search with fabrication-aware constraints. This method imposes binarization and minimum feature size constraints throughout optimization to avoid post-processing steps such as thresholding[42].

'Fabrication-aware' and robust inverse design methods embed models of fabrication variability directly into the optimization process. Even DRC-compliant devices may underperform due to fabrication-induced variability (e.g., corner rounding, proximity effects, line shortening, sidewall roughness)[43]. Robust inverse design aims to minimize fabrication sensitivity by accounting for fabrication variations throughout the design process[38]. A conventional approach to robustness involves optimizing a structure across an ideal, over-etched, and under-etched geometries[38,40]. This approach, however, assumes that all three design fields share the same topology, which may not hold for small features. Therefore actual fabrication outcomes may vary significantly from any of the three geometries[40,43]. By instead calculating the figure of merit based on lithography-predicted geometry, the parameter space can be transformed to be fabrication-aware[43].

Deep learning models are emerging as a strategy to improve robustness. For example, a deep learning model was employed to predict and compensate for fabrication deviations (Fig. 2l), resulting in significant improvements in predicting fabrication outcomes of complex features relative to uniform bias models[40].

The advancement of robust, fabrication-aware methodologies in academic research has facilitated the translation of inverse design into industry, with an emphasis on compactness and reliability. For example, Google demonstrated a silicon 4-channel CWDM demultiplexer with consistent performance across 34 chips[49] (Fig. 2r). Industry groups are also leveraging open-source platforms that were developed by academic researchers. Corning used SPINS[2] to design a compact dual-polarization silicon coupler that addresses 14 fiber modes simultaneously[50] (Fig. 2m), while Samsung used the FDTD-z platform[3] to design reliable silicon devices to explore the physical limits of photonic components[51].

Given the scaling of inverse design device footprint due to advancements in electromagnetic solvers and the extension of inverse design to commercial foundry processes, some of the first inverse design devices have undergone significant enhancements. One exemplary metric for this progress over the past decade is silicon wavelength de-multiplexer (WDM) performance. In Table 1, we present the evolution of inverse design WDMs, using the first demonstration in 2015 as a baseline. Moreover, scalable foundry-compatible inverse design enables high-performance silicon chips with system-level applications such as optical interconnects. For example, 1.12 Tb/s natively error free data transmission was achieved by combining inverse-designed WDMS, MDMs, and surface-normal couplers into a multi-dimensional communication scheme[52] (Fig. 2d).

Table 1 | **Progress in inverse design silicon wavelength de-multiplexer performance**

| Year | Channel spacing | Insertion loss | Crosstalk | No. of channels | Footprint | Fabrication process | Reference |
|---|---|---|---|---|---|---|---|
| 2015 | 250 nm | 1.8 - 2.4 dB | -11 dB | 2 | 2.8 x 2.8 $\mu m^2$ | University | Piggott et al.[53] |
| 2018 | 40 nm | 2.29 - 2.82 dB | -10.7 dB | 3 | 5.5 μm x 4.5 μm | University | Su et al.[54] |
| 2024 | 20 nm | 2 - 3.3 dB | -19 - -26 dB | 4 | 14 μm x 16 μm | Commercial Foundry | Cheung et al.[49] |
| 2025 | 15 nm | 1.86 - 3.9 dB | -41 dB | 2 | 12 x 12 $\mu m^2$ | University | Mason et al.[55] |

## Inverse design beyond silicon photonics and optical wavelengths

The inverse design process is material- and wavelength-agnostic, and parameters and fabrication constraints for any materials and wavelengths can be incorporated into the design process. Consequently, since the initial demonstrations of inverse design in silicon photonics[53,54,56,57], Ⅲ-Ⅴ materials[58,59], and photovoltaic systems[60,61], inverse design has expanded into diverse material and wavelength systems providing unique functionalities. Here we describe recent progress on inverse design in now mainstream photonic materials - silicon nitride and lithium niobate, as well as leading quantum photonic materials - diamond and silicon carbide. In this section we focus on linear inverse design, while in the next section we describe direct nonlinear inverse design in these materials. Moreover, we also describe recent progress on utilizing inverse design in terahertz and microwave wavelength ranges.

Recently, silicon nitride has emerged as a promising material in integrated photonics featuring ultra-low material loss[62] which provides dB/m order loss in telecommunication and visible wavelengths, and provides favorable nonlinear optical properties based on its third order ($\chi^{(3)}$) nonlinearities.

Similar to silicon photonics as described in the previous section, inverse design was used to achieve silicon nitride photonic blocks with better performance and compact footprints: WDM or MDMs[63,64] (Fig. 2q), power splitters[65,66] and polarization beam splitters[63,67]. Shrinking footprint is a grand challenge here because of the low index contrast between silicon nitride and the cladding which requires a larger device footprint than typical silicon photonics systems[68]. For instance, the inverse-designed WDMs can achieve low insertion loss and crosstalk with a moderate minimum feature size of 160 nm and only a 24 μm x 24 μm size device, which is significantly smaller than traditional MZI or ring resonator-based components[68,69].

In the past decade, thin-film lithium niobate (TFLN) has emerged as a photonics platform that provides low propagation loss and large $\chi^{(2)}$, $\chi^{(3)}$ nonlinearity in a broad wavelength range - from visible to terahertz wavelengths. However, similar to silicon nitride, LN (lithium niobate) has a low refractive index making the design problem challenging. Moreover, LN fabrication poses unique challenges, such as sloped sidewalls, and the LN birefringence further complicates the design process. Recently, inverse design techniques incorporating both birefringence and sidewall features have been proposed and demonstrated. Qiao et al.[70] proposed a fabrication-aware framework tailored to LNOI platforms, involving a slope-processing algorithm that decomposes sloped sidewalls into discrete layers.

One exemplary challenge in the TFLN platform is a grating coupler, because of  the low index contrast between lithium niobate and the cladding layer. Xue et al.[71] and Zhan et al.[72] show high-efficiency inverse designed grating couplers with 3.18 dB and 1.97 dB of coupling efficiency, respectively. Both studies incorporate the sidewall angle of LN into their simulation model being used in optimization, and hence the resulting simulated efficiency is closer to the actual value. However, these studies are mostly optimizing a small set of discrete parameters that describe the shape of gratings (namely, widths and gaps of gratings) and not really conducting a full scale optimization. A more general inverse-design technique for lithium niobate based on topology optimization was proposed by Shang et al.[73]. This study follows a similar design strategy as topology optimization based inverse design approaches in conventional silicon photonics, consisting of an optimization step with a continuous permittivity distribution, optimization with binarization and discrete optimization that enforces a minimum feature size requirement. After the discrete optimization step, additional optimization steps add a sidewall projection model into the simulation model to get a final design with full consideration of sidewall features. The resulting mode multiplexer (Fig. 2j), waveguide crossing, waveguide bend design gave small insertion loss and crosstalk and significant footprint reduction compared to the conventional designs. This study was demonstrated on z-cut LN, which is convenient for inverse-design in a TE mode configuration. However, TFLN devices utilizing $\chi^{(2)}$ or Pockels effect ($r_{33}$) uses x-cut LN, which requires birefringence consideration in the optimization step.

Recently, Lyu et al.[74] proposed an improved topology optimization method with both sidewall and birefringence considerations, which led to a Y-splitter design in x-cut LN with 0.70 dB insertion loss. Aside from the typical component designs for compact, high-performance TFLN photonic design, inverse design is being utilized in various functions such as mode converters for second harmonic generation and spontaneous parametric down conversion[75] and on-chip mirrors for enhanced optical gain in erbium-doped LN waveguides[76].

Performance requirements are even more stringent in the quantum regime than in classical photonics, as loss of even individual photons has dramatic consequences on the system performance. Often, quantum photonic devices operate in extreme conditions, such as low temperature, high vacuum, thereby requiring new ways of probing the systems, where inverse design plays an essential role in finding new, more suitable geometries. Moreover, many suitable quantum materials (such as diamond) have even more stringent fabrication constraints than silicon nitride or lithium niobate.

In recent years, direct patterning of inverse-designed coupler structures has been adopted to provide an efficient and simple way of light coupling for such quantum systems. For example, efficient inverse-designed power splitters and vertical couplers that can provide 25% coupling efficiency with 1 μm x 1 μm footprint in diamond were demonstrated, and they were compatible with stringent diamond fabrication[77] (Fig. 2b).  Heterogeneous structures comprising of bulk diamond hosting NV color center and an inverse-designed Gallium Phosphide (GaP) pattern on top were also demonstrated[78], with 14 fold enhancement of zero phonon line (ZPL) emission from NV center placed 100 nm below the surface. Another example is hexagonal boron nitride (hBN), which is frequently used as an insulating capping layer or itself as a host of quantum emitters. Inverse-design provides a huge advantage in footprint which is crucial in two-dimensional material systems such as hBN due to the small size of flakes. Monolithic hBN inverse designed grating coupler with 35% outcoupling efficiency from single photon emitters in hBN was demonstrated[79], as well as more advanced nanophotonic components on hBN encapsulating $MoSe_2$[80], including compact grating coupler with 11% efficiency, waveguide mirrors with 99.5% reflectance, and metasurfaces to increase the collection efficiency of dark excitons.

Silicon Carbide is yet another example of an emerging quantum material, and inverse design has been employed to design grating couplers in 4H-silicon-carbide for first harmonic (1555 nm) and second harmonic (777.5 nm), enabling a second-order frequency conversion in microring resonators[81]. Additionally, Kerr nonlinearity mediated OPO and frequency combs in 4H-silicon-carbide on insulator systems have been demonstrated, with inverse designed vertical couplers with 31% peak efficiency and broad bandwidth (1450 ~ 1600 nm)[82] . Direct inverse-design of cavity was also adopted for quantum frequency combs in Yang et al.[83], where an inverse designed Fabry-Perot cavity was used to create frequency combs in C-band and visible wavelengths through the $\chi^{(2)}$ and $\chi^{(3)}$ of silicon carbide (Fig. 2i).

Fundamentally, any system that is described by Maxwell's equations can benefit from a similar inverse design tool that is used for visible and telecommunication wavelength regimes as previously described. Terahertz integrated photonics[84,85], is an emerging field where photonics-based devices can process THz signals giving much more compact and versatile designs than those based on microwave engineering. A three-channel WDM operating at the 2 ~ 3 THz range, creating three terahertz emission antennas with different frequency ranges split from a shared co-integrated quantum cascade laser (QCL) source was demonstrated recently[86] (Fig. 2s). The same inverse design tool that is used for inverse design at telecommunication wavelengths[2], are utilized for this THz work which shows a universality of inverse design methodology. Additionally, the same team demonstrated a low-reflectivity end facet of the metallic waveguide used in the QCL device[87] and enhanced the slope efficiency (slope between power and injected current) by seven times.

For free-space terahertz photonic elements, inverse-design has been used to design a EIT (electromagnetically induced transparency) in a metasurface structure comprising of germanium (Ge) on quartz and patterned gold (Au) layer on top[88]. The proposed metasurface can modulate the EIT based on the visible light pumping into Ge film that controls the conductivity of Ge film due the carrier excitation. Electromagnetic systems with wavelengths longer than terahertz, namely, microwave and magnonic systems, are the systems where inverse design has been utilized to produce high-performance, versatile devices. S-matrix engineering has enabled parallel computing[89], filter and divider design[90], and a dual polarization surface launcher for a dielectric waveguide[91], demonstrating complex functionalities that couldn't be implemented in traditional microwave engineering. In magnonic systems, the propagation of spin waves in magnetic materials like YIG (yttrium iron garnet) can be manipulated by putting nano-patterns or applying an external magnetic field. Wang et al.[92] shows the general idea of inverse design in a magnonic device that consists of a 10 x 10 array of 100 nm x 100 nm YIG patches (represented by 10 x 10 array of magnetization). These patches can be inverse-designed for frequency multiplexing and as circulators with fixed external fields and a nonlinear switch that can be reconfigured by changing the external magnetic fields. Recently, Zenbaa et al.[93] devised a reconfigurable magnonic device that can tune the response based on 49 current sources array placed on a YIG film, providing a general scheme to implement reconfigurable magnonic devices with target transmission functionality.

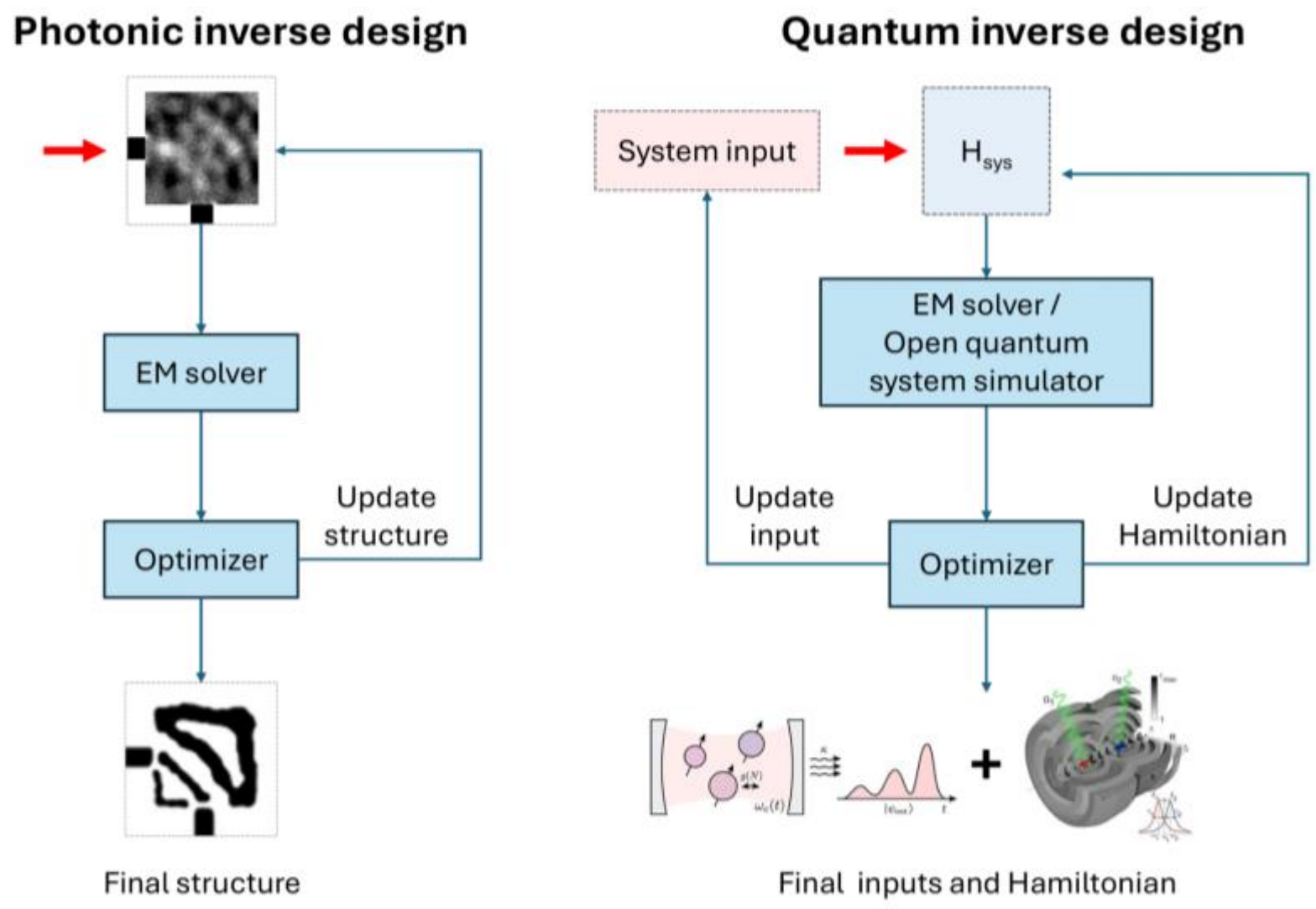

**Figure 4. Quantum inverse design**
Analogous to photonic inverse design (left), a similar approach can be applied to designing quantum systems (right). Parameters of the system (dielectric constant distribution, global electric field, optical drive modulation, etc.) are optimized to achieve the desired system Hamiltonian. In addition to electromagnetic solvers, solvers such as open quantum system simulators can be used in the inverse design loop.

Insets reproduced with permission from references [115,117]

Optical cavities (resonators) are the workhorses of laser physics, nonlinear optics, and quantum optics (cavity QED). The name of the game in laser design, optimization of nonlinear processes, or design of cavity QED systems is increasing cavity Q factor while reducing mode volume V. This leads to reduction of lasing threshold and improved efficiency, as well as threshold of nonlinear local processes, and crossing into strong coupling regime of cavity QED.

Since the early work on direct inverse design of photonic cavities to maximize Q/V[94], more recent effort has been made in this direction[95–98]. This includes inverse design of nanoscale optical resonators in photonic material platforms with dipoles approximating quantum emitters[99]. Such optimized photonic cavities for Q/V maximization have been experimentally demonstrated as well[95]. Aside from maximizing Q/V, inverse designing cavities for implementing diverse laser polarization and beam shape[100] have also been demonstrated.

Likewise, a majority of nonlinear photonic circuits utilize high-Q resonators to get resonance-enhanced nonlinear effects. However, such nonlinear effects are heavily affected by the dispersion of the resonators. Unlike traditional dispersion engineering methods which mostly optimize waveguide cross sectional geometry[101], inverse-design enables more robust and versatile dispersion engineering[102]. Prior work on inverse design of nonlinear optical systems has been focused on inverse design of coupled linear problems (eg. Yang et al.[83]), but recent efforts have emerged where nonlinear equations are directly optimized to achieve target performance.

In the application of Kerr-comb generation, two complementary efforts illustrate the approach. First, "meta-dispersion" treats mode-by-mode frequency splitting as a programmable degree of freedom, letting designers match user-specified comb spectra and flatten bands via controlled multimode hybridization—effectively inverting the Lugiato–Lefever dynamics to a spectral target[103].

In Zhang et al.[104], the authors used a genetic algorithm to optimize the parameters of concentric microresonators in conjunction with a DNN-based LLE equation solver to enhance the flatness of soliton combs in silicon nitride. Similarly, in Lucas et al.[103], the shape of the tantala photonic crystal ring resonator was optimized by creating 'meta-dispersion' that enables target comb shapes such as flat-top or gaussian-shape combs. Alternative approach for dispersion engineering of microresonator is inserting an inverse-designed component in a resonator with tailored phase response. In silicon photonics, such functionality was experimentally demonstrated[102] where a reflective element with target frequency-dependent reflection phase is placed in a microresonator. Recent demonstrations of inverse-designed reflectors in silicon nitride[64,105] successfully demonstrated high reflectivity (97 ~ 98%) in a broad optical wavelength range, which is promising for realizing high-Q on-chip Fabry-Perot resonators with dispersion engineering. Another approach is inverse designing a mode-converter between different mode families to design a target mode-crossing[106]. Finally, comb initiation and stabilization can be cast as an optimization problem over pump parameters. Towards this goal, genetic-algorithms have been employed to discover operating points for stable, target-shaped combs[107].

Alongside Q-factor and dispersion engineering in nonlinear integrated photonics, inverse design has been employed to directly optimize nonlinear interactions. In Jia et al.[108], phase-matching for four-wave mixing was incorporated as the loss function for a silicon-on-insulator grating cavity at telecom wavelengths, yielding a compact structure with high-Q resonances at target wavelengths, and enhanced pair-generation rate and coincidence-to-accidental ratio. Inverse design has also been applied to nonlinear metasurfaces, particularly those based on multiple quantum wells (MQWs)[109]. MQWs supporting intersubband transitions (ISTs) exhibit giant nonlinearities that can be engineered via heterostructure thickness optimization at the fundamental and harmonic wavelengths. Combined with resonator geometry optimization to excite Mie modes that enhance pump coupling to IST and local field enhancement, large SHG efficiencies have been demonstrated[110]. Recently, a topology-optimized MQW metasurface based on optimization of IST heterostructure and far field directionality with fabrication constraint consideration was experimentally demonstrated with a record SHG efficiency of 0.38 $mW/W^2$[111]. For plasmonic platforms, a $TiO_2$-$Al_2O_3$-$HfO_2$ nanolaminate with broken z-symmetry was combined with deep learning-based inverse design metallic structures, yielding enhanced SHG response at 840nm excitation[112].

Similar to photonic inverse design, inverse design of quantum systems implies posing a quantum objective (e.g., a target steady-state entangled density matrix) and optimizing the electromagnetic environment, or drive of the quantum system so that this target state is achieved (Figure 4). Beyond linear photonics optimization in quantum materials, as described in the previous section, or optimization of resonators or multi-port interferometer[113] for quantum optics experiments, previous efforts have also focused on the space of quantum control and generalized Floquet driving. This line of "quantum optimization" was initially pursued using gradient-based ascent methods, in which the objective function is defined by specific metrics of the quantum systems[114]. Recently, these prior methods led to utilizing a more robust numerical optimization landscape. For example, adjoint optimization has been used to find an optimal drive for a quantum system, which leads to superradiant effects or improved quantum transduction efficiency in inhomogeneously broadened ensembles[115,116]. Recent efforts on topology-optimized cavities have moved beyond quantum control problems: under continuous-wave drive and using moderate-index dielectrics, the design explicitly maximizes dissipative coupling and prepares high-fidelity Bell and W states of spatially separated emitters, linking photonic structure directly to open-system steady states[117] (Fig. 2n). In parallel, topological waveguide-QED theory clarifies which bath properties are worth targeting: band gaps, edge modes, and nontrivial winding can host chiral bound states that mediate long-range, tunable interactions and topology-dependent sub/superradiance[118]. For quantum light sources, multi-objective formulations that jointly maximize Purcell factor and collection efficiency already point toward end-to-end objectives (brightness, indistinguishability) under realistic geometry and fabrication constraints[119]. A practical recipe follows: encode fidelity or source metrics in the objective; co-optimize coherent/dissipative couplings by shaping the local density of states; and include tolerances, spectral-diffusion, and bandwidth terms so designs transfer cleanly to a variety of suitable quantum materials with reproducible, high-fidelity performance.

## Conclusion

The field of photonic inverse design has been advancing rapidly over the past decade, moving beyond proof of concept demonstrations relying on university fabrication facilities. The use of inverse design in photonics is certainly not only an academic curiosity anymore, but is getting a broad acceptance beyond academia, where it is actively employed to address problems where traditional photonics fails (such as higher efficiency, broadband couplers for optical interconnects made in commercial foundries[120]). Major progress has been made in translating inverse design to commercial silicon foundries, and in developing new electromagnetic solvers and tools capable of addressing larger design footprints, even full metasurface design. Being materials and wavelength agnostic, inverse design has naturally moved into materials beyond silicon and wavelengths outside of the optical band. Recognizing the success of photonics inverse design, similar approaches are emerging in design of scalable nonlinear and quantum systems, in the context of dispersion engineering, quantum control, as well as design of environments to achieve target hamiltonians.

Looking forward, we will certainly see even more inverse design applications to other research areas and fields, as this approach has proven itself as a tool to solve difficult problems where human intuition and standard building blocks fail. In fact, big initiatives are already emerging in academia, government, as well as in industry focused on the use of AI tools in scientific discovery and to accelerate engineering - such as the discovery of new materials, chip design, or to advance quantum information hardware[121–123]. In large-scale photonics inverse design, we anticipate that AI tools may be helpful for structure initialization or defining the objective function, which are now still dependent on the experienced researcher. Further advancement of computational bounds (in particular making them stricter) will eventually help choose initial device footprint for a specific problem instead of going through trial and error[124–127]. And with further AI hardware advancements, we expect that even more powerful electromagnetic simulators and inverse design tools will emerge[3,5], capable of solving larger device footprints or possibly designing full systems.

While inverse design has already proven itself as a tool that can greatly enhance research and engineering capabilities, it will not be a replacement for skilled researchers and their creativity. It is important to remember that although these tools seem very powerful on the outside, they are not intelligent or creative; instead, they are only performing complex computational tasks very fast - the tasks that we as researchers programmed them and assigned them to do. In the case of photonics, these are essentially Maxwell's equations solvers combined with adjoint optimization/backpropagation techniques, allowing fast calculation of gradients and gradient descent. Without human ingenuity, critical thinking, and guidance in defining problems and bringing up new ideas, even the best computational hardware and software will not be capable of making real breakthroughs in science and advancing knowledge. Therefore, the availability of these powerful tools should not be the reason to stop teaching students underlying subjects such as electromagnetism or photonics. On the contrary, the success of inverse design should be the motivation to strengthen education and training across subjects (such as electromagnetics + computer science), as this is an example of how creative researchers who are well trained in multiple fields can greatly advance science and engineering.

## Acknowledgments

We acknowledge financial support from the US Department of Energy ESTEEM Center (SM, GHA), AFOSR under award no. FA9550-23-1-0248 (GHA, LS), Advanced Micro Devices through SystemX (SM), Samsung (SE and GHA), and Marvell through SystemX (SE). We thank our former team member Logan Su for inverse designs of waveguide bends used in Figure 1. SE acknowledges support from the Shoucheng Zhang Graduate Fellowship and Korea Foundation for Advanced Studies Overseas Ph.D. Fellowship.

## Competing interests

JV is a cofounder and a scientific advisor of Spins Photonics Inc. SM, GHA, and JV are inventors on patent application no. PCT/US2025/036120 titled "Wavelength Division Component Integrated with Distributed Bragg Gratings through Co-optimization". SE, GHA, and JV are inventors on patent application no. PCT/US2025/036482 titled "Single-Mode Polarization Insensitive Grating Coupler".